\documentclass[10pt]{article}
\usepackage{amsmath,amsfonts,amsthm,amssymb,amscd}
\numberwithin{equation}{section}

\newcommand{\be}{\begin{equation}}
\newcommand{\ee}{\end{equation}}
\newcommand{\bs}{\begin{split}}
\newcommand{\es}{\end{split}}
\newcommand{\ba}{\begin{align}}
\newcommand{\ea}{\end{align}}
\newcommand{\basl}[1]{\begin{align}\begin{split}\label{#1}}
\newcommand{\bas}{\begin{align}\begin{split}}

\newtheorem{theo}{Theorem}[section]
\newtheorem{prop}[theo]{Proposition}
\newtheorem{lemm}[theo]{Lemma}

\newtheorem{defi}[theo]{Definition}

\newcommand\fpr{\hfill$\Box$\null}

\newcommand\N{\mathbb{N}}

\newcommand\R{\mathbb{R}}

\title{Compositions of states and observables in Fock spaces}

\author{ L. Amour, L. Jager, J. Nourrigat}

\date{Universit\'e de Reims, France}

\begin{document}

\maketitle

\begin{abstract}
\noindent
This article is concerned with compositions in the context of three standard quantizations in the Fock spaces framework, namely, anti-Wick, Wick and Weyl quantizations. The first one is a composition of states and is closely related to the standard scattering identification operator encountered in Quantum Electrodynamics for time dynamics issues (see \cite{HS-RD}, \cite{DG}). Anti-Wick quantization and Segal-Bargmann transforms are implied here for that purpose. The other compositions are  for observables (operators in some specific classes) for the Wick and Weyl symbols. For the Wick symbol of the composition of two operators, we obtain an absolutely converging series and for the Weyl symbol, the remainder term of the expansion is absolutely converging, still in the Fock spaces framework.
\end{abstract}

\parindent=0pt

\


\


\tableofcontents

\parindent=0pt
\parindent = 0 cm

\parskip 10pt
\baselineskip 12pt

\section{Introduction.}\label{s1}

The purpose of this article is to present within the Fock space framework three different results concerning three common quantizations, that is, anti-Wick, Wick  and Weyl quantizations.

In the first part of this work, we are concerned with the anti-Wick quantization together with the operator commonly called the scattering identification operator for some spin boson model (see below).  Namely, our first aim is to show that  the scattering identification  can also be defined using the anti-Wick quantization.

For any infinite dimensional real separable Hilbert space $H$,  ${\cal F}_s (H_{\bf C} )$ denotes the symmetrized Fock space over $H_{\bf C}$ where $H_{\bf C}$ stands for the complexification of $H$. (See \cite{RSII}, vol. I, sect II.4.)  
For all $X$ in $H$,  $a(X)$ and $a^{\star}(X)$ denote annihilation and
creation operators (unbounded operators in  ${\cal F}_s (H_{\bf C} ))$ associated with  $X$. 
(See \cite{RSII}, vol. II, sect X.7.)  
The state $\Psi_0$ stands for the vacuum. The subspace  ${\cal F}_s^{\rm fin} (H_{\bf C} )$ denotes  the set  of all elements written as finite linear combinations of  $a^{\star}(X_1)\dots a^{\star}(X_m)\Psi_0$
 with $X_j$ in $H$.  For some special Hilbert space $H$, the  identification operator is an unbounded operator in
   ${\cal F}_s (H_{\bf C} )$ defined with the help of some internal composition law in ${\cal F}_s ^{\rm fin} (H_{\bf C} )$ introduced in  \cite{HS-RD}. See also  \cite{DG}  and \cite{SP}. This internal composition law can be 
   defined for any Hilbert space $H$. 
Let us recalled below the definition as it is written in  \cite{SP} in formula (17.92). See also  \cite{DG}.

      \begin{defi}\label{def-prod} One denotes by  $I$ the bilinear map from
       ${\cal F}_s ^{\rm fin} (H_{\bf C} ) \times {\cal F}_s  ^{\rm fin} (H_{\bf C} )  $ to ${\cal F}_s (H_{\bf C} )$
     satisfying for all  $X_1,\dots,X_m$ and all $Y_1$, ... $Y_p$ in $H$:
     \be\label{equ-prod}  I \Big ( a^{\star}(X_1)\dots a^{\star}(X_m)\Psi_0 \  , \  a^{\star}(Y_1)\dots a^{\star}(Y_p)\Psi_0
      \Big ) =  a^{\star}(X_1)\dots a^{\star}(X_m)a^{\star}(Y_1)\dots a^{\star}(Y_p)\Psi_0. \ee
     \end{defi}

       The above  product is defined for  $f$ and $g$  in
       ${\cal F}_s ^{\rm fin} (H_{\bf C} )$ and we shall see in Proposition \ref{norme-prod-C} that its domain of  definition can be extended. Also recall here that creation operators are commuting.

        In Section \ref{s2-5}, we study the connection between this composition law $I$
       and the  Segal Bargmann transform. Preliminary,
        we present  in Section \ref{s2-2} three variants of the   Segal Bargmann transform of $f$ (Definition \ref{trois-TF}), for all $f$ in
        ${\cal F}_s ({H}_{\bf C} )$. The first one, denoted $T^{FF} f$,
        is an element of the
        Fock space ${\cal F}_s ({H}_{\bf C} ^2)$. The two others are depending on a (semiclassical) parameter $h>0$. The second one, denoted  by $T_h^{FH} f$, is a function on  $H^2$ being Gateaux anti-holomorphic if identifying $(q , p)$
        with $q+ip$. The third one is denoted by $T_h^{FW} f$ and is a  $L^2$ function on some space denoted  $B^2$ equipped with a Gaussian measure $\mu _{B^2, h}$. The properties of  the space $B$ and of the 
        measures $\mu _{B, h}$ on $B$ and  $\mu _{B^2, h}$ on $B^2$ are recalled in Section \ref{s2-1}. One may say that  spaces $B$ suitable for this construction are Wiener extensions of $H$.

We prove in Theorem \ref{prod-SB}   that, when $f$ and $g$ are in some suitable domain included in
        ${\cal F}_s ({ H}_{\bf C} )$:
\be\label{prod-I-SBH}  T_h^{FH}  I(f , g) (q, p) =  T_h^{FH}f (q, p) \ T_h^{FH}g (q, p)  \ \ \ \ (q , p)\in H^2 \ee
       \be\label{prod-I-SBW} T_h^{FW}  I(f , g) (q, p) =  T_h^{FW}f (q, p) \ T_h^{FW}g (q, p)
         \ \ \ \ a.e. \ (q , p)\in B^2. \ee
Therefore, via  Segal Bargmann transform, the composition law  $I$ is reduced to nothing else than the usual multiplication for functions.

We are now extending the domain of definition of the law  $I$.

For any $R\geq 1$, denote by ${\cal F}_s^{R} (H_{\bf C})$ the subspace of all $f = (f_n)$ in
${\cal F}_s (H_{\bf C})$ such that the following series is converging:
\be\label{norme-FsR} \Vert f \Vert _R ^2 =  \sum _{n\geq 0} R^n \Vert f_n\Vert ^2 < \infty. \ee
In particular, ${\cal F}^{1}_s (H_{\bf C})= {\cal F}_s (H_{\bf C}) $.

\begin{prop}\label{norme-prod-C}  For all $R$, $R'$ and $R''$ satisfying $R \geq 1$,
  $R'\geq 1$, $R''\geq 1$ and  $(1/R'') = (1/R) + (1/R')$, the  composition law $I$  in Definition \ref{def-prod} is extended to a continuous bilinear map from
  ${\cal F}_s^{R} (H_{\bf C})   \times {\cal F}_s^{R'} (H_{\bf C})  $  into
  ${\cal F}_s^{R''} (H_{\bf C}) $. For all $f$ in ${\cal F}_s^{R} (H_{\bf C})$
  and  $g$ in  ${\cal F}_s^{R'} (H_{\bf C})$, we have:
  \be\label{norme-prod-2} \Vert I ( f , g)  \Vert _{R''}
  \leq  \  \Vert f  \Vert _{R}  \  \Vert g\Vert _{R'}.  \ee

  \end{prop}
  This is proved is Section \ref{s2-5}

We now turn to the scattering  identification operator itself. First, let us give the form of this 
operator, and we shall give more details below. In this theory, we have an infinite dimensional 
Hilbert space $H$ and a finite dimensional Hilbert space ${\cal H}_{sp}$ describing the spins 
of the particles. Then we are interested in some element $U$ of ${\cal F}_s ({H} _{\bf C}) \otimes {\cal H} _{sp}$. 
More precisely, for each $R>1$, $U$ is in  ${\cal F}_s^R ({H} _{\bf C}) \otimes {\cal H} _{sp}$.  
One now chooses a Hermitian basis
 $S^{(\lambda)}$  $(1\leq \lambda \leq d)$ of  ${\cal H} _{sp}$. We denote by $U_{\lambda}$
the components of the ground state $U$ in  this  basis,
\be\label{basis} U = \sum _{\lambda = 1}^d  U_{\lambda}\otimes  S^{(\lambda)}. \ee
The scattering identification $J$ is defined for all $\varphi $
in  ${\cal F}_s^{R} (H_{\bf C})$ (with $R>1$) by:
\be\label{def-J-lambda} J \varphi = \sum _{\lambda = 1}^d J _{\lambda }  \varphi \otimes  S^{(\lambda)}
\hskip 2cm  J _{\lambda }  \varphi = I  (U_{\lambda} , \varphi ).  \ee
By Proposition \ref{norme-prod-C}, $J$ is bounded from
 ${\cal F}_s^{R} (H_{\bf C})$ into   ${\cal F}_s(H_{\bf C})\otimes {\cal H} _{sp}$ for all $R>1$.

       In some spin boson models, an important role is played by   some  Hamiltonian acting 
       in the Hilbert space ${\cal F}_s ({H} _{\bf C}) \otimes {\cal H} _{sp}$
   where  ${ H}$ is the configuration space for the one photon particle states, that is to say,
   the space of all $f$ in  $L^2(\R^3 , \R^3)$ satisfying $k\cdot f(k) = 0$ almost everywhere for $k\in\R^3$,
    and ${\cal H} _{sp}$ is some finite dimensional space. Some  Hamiltonians of this type are 
      defined in \cite{S-1989}, \cite{HS-RD}, \cite{C-G}, ...  For  a model related to NMR, see \cite{ALN} or  \cite{Reu} in a more physical viewpoint.
One shows (see \cite{C-G} or  \cite{S-1989}) that for  this Hamiltonian, there exists an eigenfunction $U$ with unit norm,  (this fact comes from \cite{H} or \cite{S-1989}), with an eigenvalue  located at the bottom of the spectrum of the Hamiltonian. 
If some coupling  constant appearing in the operator is small enough, this ground state is unique
   up to a constant multiplicative factor. One also shows in \cite{HS-RD} that   $U$ belongs to  ${\cal F}_s^{R} (H_{\bf C})$ for every $R\geq 1$. Perhaps these techniques could be applied in other situations.

Our next purpose is to show that the operators $J _{\lambda } $ are defined  by  the
anti-Wick quantization. First, let us recall the definition. 

     With some bounded functions  $F$ on $H^2$, satisfying suitable conditions, we one can usually 
     define an anti-Wick operator
 $Op_h ^{AW}  (F)$, bounded on   ${\cal F}_s (H_{\bf C})$ and depending of  $h>0$ in the following way. 
 First, we choose a Wiener extension $B$ of  $H$. See Theorem \ref{t1.1} below  for definition and 
   existence. For each $h>0$, this space is provided with a  Gaussian measure $\mu _{B , h}$ 
   with  variance $h$, and $B^2$ is provided with a  Gaussian measure $\mu _{B^2 , h}$. 
   The definition of the anti-Wick quantization uses one of the three variants of the 
   Segal transformation of Definition \ref{trois-TF}, namely the transformation $T_h^{FW}$. 
   If $f$ is an element of ${\cal F}_s (H_{\bf C})$, $T_h ^{FW}f$ is an element of 
   $L^2 (B^2 , \mu _{B^2,h})$. If $F$  is a bounded function on $H^2$,  we can define a related 
   anti-Wick operator only if   $F$ has  a stochastic extension  $\widetilde F$, in the sense of Definition \ref{def-ext-stoc}
below,  in $L^2 (B^2 , \mu _{B^2,h})$. If  $F$ is bounded then the stochastic extension $\widetilde F$ is also bounded.  Then, the  operator  $Op_h ^{AW}  (F)$ is the unique operator such that, for each $f$ and $g$ in ${\cal F}_s (H_{\bf C})$:
\be\label{OPAW}  < Op_h ^{AW}  (F) f , g> = \int _{B^2} \widetilde F (x) T_h^{FW} f (x) \
\overline { T_h^{FW} g (x) } d\mu _{B^2 , h} (x) \ee

The main result of the first part of this article, proved in Section \ref{s2-5}, is the following one 
(with $h=1$). In the statement, one uses another variant of the Segal Bargmann transformation, 
denoted by $T_h^{FH}$, also defined in  Definition \ref{trois-TF}. For each $U$ in 
${\cal F}_s (H_{\bf C})$, $T_h ^{FH}U$ is a regular function on $H^2$. It has a stochastic extension, 
which is  $T_h ^{FW}U$ (Proposition \ref{2-transfo}). 

\begin{theo}\label{t-J}
If $U$ is of the form (\ref{basis}), with $U_{\lambda }$ in ${\cal F}_s^{R} (H_{\bf C})$ for all  $R>1$,
  one has:
\be\label{J-AW}  J_{\lambda } = Op_1 ^{AW}  (F_{\lambda }),\quad{\rm with}\
F_{\lambda } = T_1^{FH} U_{\lambda }. \ee
\end{theo}

The  specificity of  this operator  is that  $T_1^{FH} U_{\lambda }$
is not  bounded.  However, the integral (\ref{OPAW}) makes sense if $f$ is 
in ${\cal F}_s^{R} (H_{\bf C})$ for some $R>1$, because the product 
$\widetilde F_{\lambda}  (x) T_h^{FW} f (x) $ is the image by $T_h^{FW}$ 
of  $I(U_{\lambda } , f)$  (Theorem \ref{prod-SB}), and $I((U_{\lambda } , f)$ is 
a well defined element of ${\cal F}_s (H_{\bf C})$ by Proposition \ref{norme-prod-C}.   
Therefore the  operator   $J_{\lambda }$
defined  in that way is unbounded:  its domain contains the  union of the   ${\cal F}_s^{R} (H_{\bf C})$.

There are two other  ways to make a relation between an operator and a   function $F$ on $H^2$. 
the Weyl and the Wick symbols. Coherent states $\Psi_{X h} $ are  elements  of  the Fock space
${\cal F}_s (H_{\bf C})$ indexed by  $X = (q , p)\in H^2$ and depending on  $h>0$, defined  by:
\be\label{C-S} \Psi_{X, h }  = \sum _{m\geq 0}
 {e^{-{|q|^2+ |p|^2  \over 4h }} \over (2h)^{m/2} \sqrt {m!} } (q+ip) \otimes \cdots \otimes (q+ip).\ee
We remind  that we call Wick symbol of an operator  $A$ in  ${\cal F}_s (H_{\bf C})$ with a domain containing the
 coherent states $\Psi_{X , h}$
$(X\in H^2)$,  the  following function defined on  $H^2$ by (the scalar product is linear with respect to the first variable):
\be\label{S-Wick}  \sigma _h^{wick} (A) = < A \Psi_{X , h} , \Psi_{X , h} >. \ee 
For the Weyl symbol, see \cite{A-J-N-1} and \cite{A-J-N-2}). 
Let us  denote by 
 $\sigma _h {Weyl} (A)$ and $\sigma _h {Wick} (A)$ the  symbols 
of an  operator $A= Op _h^{AW} (F)$, for a suitable  given function $F$ on $H^2$. 
When $F$ is a suitable function, (see below), let $Op_h^{weyl} (F)$ be the 
correspondinng Weyl operator. 
 The relation between these three functions is given by the heat operator.

    The heat operator is defined for each
measurable bounded function $F$ in  $H^2$ admitting a stochastic extension
$\widetilde  F$ in $L^1 (B^2 ,\mu_{B , h})$ (see Definition \ref{def-ext-stoc})
 and for each $h>0$  by:
\be\label{Ht}(H_{h} F) (X) = \int _{B^2} \widetilde  F (X+ Y)d\mu_{B^2,h} (Y)\ee
for $X\in H^2$.

It is proved in \cite{A-J-N-2}) (formula (2.13) that, if $A= Op_h^{AW} (F)$, where the function 
$F$ on $H^2$ must have a stochastic extension $\widetilde F$, as we saw:
$$ \sigma _h ^{Wick } (A) = H_h F$$
It is proved in \cite{A-L-N-3} (formula (33)) that, for each suitable function $G$ admittind 
a stochastic estension $\widetilde G$:
\be\label{Weyl-Wick}  \sigma _h ^{wick} \Big ( Op_h ^{weyl}(G) \Big ) (X) =  H_{h/2} G (X),
\hskip 2cm X\in H^2.  \ee
Therefore, if   $A= Op_h^{AW} (F)$, we have also:
$$   \sigma _h ^{Weyl}(A) =   H_{h/2} G $$

The function $F_{\lambda } = T_1^{FH} U_{\lambda }$  is Gateaux holomorphic (Proposition \ref{DSE}). 
Since any  holomorphic function remains fixed by this operator
then these three symbols of $J_{\lambda}$ are equal.

The product $I$ of  two coherent states, defined in (\ref{C-S})  is another formula that can be useful.
We prove the following equality  in  Section \ref{s2-5}, for all $X$ and $Y$ in  $H^2$:
 \be\label{equ-I-EC} I ( \Psi_{X, h } , \Psi_{Y, h }) = e^{ \frac {1} {2h} X\cdot Y} \Psi_{X+ Y, h }. \ee

In the second part of this article, we consider Wick symbols and Section \ref{s3} is devoted to the Wick symbol for compositions.
We remind  that we call Wick symbol of an operator  $A$ in  ${\cal F}_s (H_{\bf C})$ with a domain containing the
 coherent states $\Psi_{X , h}$
$(X\in H^2)$,  the   function defined on  $H^2$ by (\ref{S-Wick}).

In  Section \ref{s3},  we shall state and prove  a formula giving the Wick symbol of the
composed $A\circ B$ of two bounded operators  $A$ and $B$ in
${\cal F}_s(H_{\bf C})$, as the sum of an absolutely convergent series.
In finite dimension, this is a result  of  Mizrahi \cite{Mizrahi}
and Appleby \cite{Appleby}.

In the third part of this work, in  Section \ref{s4}, we study the  composition  for the  Weyl quantization.
The  Weyl calculus is one of the possibilities to define an operator  in  the Fock space ${\cal F} _s (H_{\bf C})$, depending on a parameter
$h>0$ and denoted by $Op_h^{weyl}(F)$, with a suitable
function $F$ defined in   the space $H^2$.
The simplest way to define this operator, at least implicitly, is to give its Wick symbol.
In order to define the Weyl operator $Op_h ^{weyl}(F)$ and in order for it to be bounded, it is necessary that  $F$ has stochastic extension  $\widetilde F$ in  $B^2$
        (in the sense  of  the  Definition \ref{def-ext-stoc}).  Then one  has:

This step is described with more details in
\cite{A-J-N-1} and \cite{A-L-N-3}.

We now recall a set of  functions $F$ for which this calculus can be applied.  See \cite{A-J-N-2} and \cite{J}.

\begin{defi}\label{d1.1}   For each
real separable Hilbert space $H$ and for each  nonnegative quadratic form
$Q$  on
$H^2$, let  $S (H^2, Q) $ be the class of functions $f \in C^{\infty }
(H^2)$  such that  there exists $C(f) >0$ satisfying, for any integer $m\geq
0$,
\be\label{1.1}|(d^m f ) (x) (U_1 , \dots , U_m ) | \leq C (f)\,
Q( U_1) ^{1/2} \dots  Q( U_m) ^{1/2}.
\ee
The smallest  constant satisfying
$(\ref{1.1})$ is denoted by $\Vert f \Vert _Q$.
\end{defi}

We denote by $A_Q$ the linear map in $H$ satisfying:
\be\label{1.100}Q(X) = ( A_QX ) \cdot X  \ee
with $A_Q\in {\cal L}(H^2)$, $A_Q=A_Q^*$, $A_Q$  nonnegative.
In what follows, $A_Q$ will also be trace class in
$H^2$. The idea of defining a class of symbols in this way, in  a
quantization purpose, with the help
of a quadratic form on the phase space, goes back to
H\"ormander \cite{HO} and Unterberger \cite{U}.

If   $A_Q$ is trace class then any function $F$ lying in  $S(H^2 , Q)$  has a stochastic extension
(in the sense of  Definition \ref{def-ext-stoc}), (see \cite{J}),
and one can associate with it a bounded operator  $Op_h ^{weyl}(F)$
following \cite{A-J-N-1} and \cite{A-L-N-3}. This operator   satisfies
(\ref{Weyl-Wick}). For every function $F$ belonging to  $S(H^2 , Q)$, the heat operator  can also be written as:
\be\label{heat} H_{h/2} F = \sum _{k=0}^{\infty} \frac { (h/4)^k} {k!} \Delta ^k F. \ee
The series is convergent and defines an element of  $S(H^2, Q)$.

One shows in  \cite{A-L-N-3} that, if  $F$ and $G$ are in
$S(H^2 , Q)$ and if  $A_Q$ is trace class then the composition $Op_h ^{weyl} (F) \circ  Op_h ^{weyl} (G)$
can be written as $Op_h ^{weyl}(K_h^{weyl} (F, G) )$ with $K_h^{weyl} (F, G) $ belonging to   $S(H^2,4 Q)$.
The purpose of  Section \ref{s4} (Theorem \ref{t1.6})  is to give an expression of $K_h^{weyl} (F, G) $ written as an  absolutely convergent series.

\section{Anti-Wick quantization and the scattering identification operator.}\label{s2}

\subsection{Wiener extension and Segal isomorphism. }\label{s2-1}

Let us recall how the Fock space ${\cal F}_s(H_{\bf C})$ is made isomorphic to some $L^2$ space  for any  Hilbert space $H$.

\begin{theo}\label{t1.1} (Gross \cite{G-1}-\cite{G-4}, Kuo \cite{KU}).
Let $H$ be a real separable Hilbert space. Then, there exists a (non unique)
Banach space $B$ containing $H$,
such that  $B' \subset H' = H \subset B$, each space being dense in the
following, and for all $h>0$, there exists a probability measure
$\mu_{B , h}$ on the Borel $\sigma$-algebra of  $B$ satisfying,
\be\label{1.1b} \int _B e^{i a(x) } d\mu _{B, h} (x) = e^{-{h\over 2} |a|^2 },\quad a\in B'.\ee
Here, $|a|$ denotes the norm in  $H$ and the notation $a(x)$ stands for
the duality between $B'$ and $B$.
\end{theo}

 The exact conditions to be fulfilled by
$B$ together with the properties involved in this paper are recalled
in \cite{A-J-N-1}. In the following, we say that  $B$ is a Wiener extension of
  $H$. Then, $B^2$ is a  Wiener extension of  $H^2$.

 We  recall (\cite{KU}) that,  for all $a$ in
$B'\subset H$, the mapping $B \ni  x \mapsto a(x) $ belongs to
$L^2(B , \mu _{B , h})$, with a norm equal to ${h}^\frac{1}{2} |a|$.
Thus, the mapping  associating with every $a\in B'$, the  above function
considered as an element of $L^2(B , \mu _{B , h})$,
can be extended by density to a mapping
$a \mapsto \ell_a$ from $H$ into $L^2(B , \mu _{B , h})$.

 We now  recall the Segal isomorphism from    ${\cal F}_s(H_{\bf C})$
 to $L^2 (B , \mu _{B, h})$.

One chooses a Hilbertian basis $(e_j)$ of  $H$.
  For all multi-indices $\alpha$ (a map from $\N$ into $\N$
with $\alpha_j=0$ except for a finite number of indices), one defines an element $u_{\alpha}$
 of  ${\cal F}_s (H_{\bf C})$ by:
  \be\label{u-alpha} u_{\alpha} =  (\alpha !)^{-1/2} \left [\prod _j
 \Big ( a^{\star} (e_j) \Big ) ^{\alpha _j} \right ] \Psi_0
 =  (\alpha !)^{-1/2} (a^{\star}(e) )^{\alpha } \Psi_0  \ee
 where  $\Psi_0$ is the vacuum state.  We know that  the $ u_{\alpha}$  constitute a Hilbertian basis of  ${\cal F}_s (H_{\bf C})$.
We also have:
$$ u_{\alpha} =
  (\alpha !)^{-1/2} \sqrt{|\alpha| ! }\  \Sigma  \left(
  \bigotimes_{j} e_j^{\alpha_j} \right)
  $$
 where  $\Sigma $ denotes the  orthogonal projection from the Fock space  ${\cal F}(H_{\bf C})$
 to the symmetrization  ${\cal F}_s(H_{\bf C})$.

 \begin{defi}\label{segal} One denotes by $S_h $ the continuous linear mapping
from   ${\cal F}_s^{\rm fin} (H_{\bf C})$  into   $L^2 (B , \mu _{B, h})$ satisfying
 for every multi-indices $\alpha $:
  \be\label{def-segal} S_h u_{\alpha} = \prod _{\alpha _j \not= 0} H_{\alpha _j}
 \left   ( \frac {\ell _{e_j} (x)} { \sqrt {h}}    \right ) \ee
 where  the $H_n$ are the Hermite  polynomials on  $\R$ (chosen as an orthonormal system in  $L^2 (\R , \mu _{\R , 1} $).
 \end{defi}

 One shows (see \cite{Jan}) that  $ S_h$ is extended to an unitary isomorphism  (Segal isomorphism) from
 ${\cal F}_s(H_{\bf C})$  into   $L^2 (B , \mu _{B, h})$.

\subsection{Segal Bargmann transform. }\label{s2-2}

Let $H$ be a separable real Hilbert space of infinite dimension.
Let ${\cal F}_s (H_{\bf C})$ be the associated  Fock space.
We now define three   Segal Bargmann transforms for any element in
 ${\cal F}_s (H_{\bf C})$.
 The first one is an
 element of  ${\cal F}_s (H_{\bf C} \times H_{\bf C} )$ and is denoted by  $T ^{FF}f$.
 The second one is a function on  $H^2$ and is denoted by  $T_h ^{FH}f$.
 The third transform, denoted by $T_h ^{FW}f$, is an element of
 $L^2 ( B^2 , \mu _{B^2 , h}) $
  where
 $B$  is a Wiener extension of   $H$ (see Section \ref{s2-1})
and  $\mu _{B^2 , h}$ is the Wiener measure with  variance
 $h$ on  the  Borel $\sigma-$algebra of  $B^2$.
 The connection between the two first transforms is realized, in some sense, with the stochastic extension notion. The relation between the last two transforms is carried out with the reproducing kernel (see Section \ref{s2-4}).

Coherent states (\ref{C-S}) are involved in order to define $T _h^{FH}f$. For the sake of clarity, in notations, we dissociate the notions concerning
the Fock space ${\cal F}_s (H_{\bf C} \times H_{\bf C} )$
and those for  ${\cal F}_s (H_{\bf C})$ by  using an index or an exponent $2$. In particular,  with ${\cal F}_s (H_{\bf C} \times H_{\bf C} )$, the scalar product is denoted by $< \cdot , \cdot >_2$, the functor $\Gamma$
is written as $\Gamma _2$ and $S_{h , 2} $ stands for the Segal isomorphism. Recall that it is linear with respect to the left variable.

 \begin{defi}\label{trois-TF}  Let $H$ be a real separable   space of  infinite dimension.

i) The mapping $T^{FF}$ acting from  ${\cal F}_s (H_{\bf C})$ into
${\cal F}_s (H_{\bf C} \times H_{\bf C} )$ is defined by:
\be\label{TFF} T^{FF} = \Gamma (U_-) \hskip 2cm U_{\pm} (X) = (1/\sqrt {2} ) ( X , \pm i X) \ \ \ \ X\in
H_{\bf C}. \ee
ii) $T_h^{FH}$ ($h>0)$ maps any element  $f$ in
${\cal F}_s (H_{\bf C})$ to a  function $T_h^{FH}f$ defined on  $H^2$ by:
\be\label{TFH} ( T_h^{FH}f) (X) = \frac { < f , \Psi_{X , h} > } { < \Psi_0 , \Psi_{X , h} > }
= e^{ \frac {|X|^2} {4h} }  < f , \Psi_{X , h} >.  \ee
iii) Suppose that $B$ is a Wiener extension  of  $H$. We set:
\be\label{TFW}  T_h^{FW } = S_{h, 2 }  \circ T_h^{FF} \ee
where  $S_{h, 2 }  : {\cal F}_s (H_{\bf C} \times H_{\bf C} ) \rightarrow L^2 ( B^2 , \mu _{B^2 , h})$
is the Segal isomorphism (see Definition \ref{segal}).

\end{defi}

The  second\ equality  in  (\ref{TFH}) comes from the following equality:  for all  $X$ and  $Y$ in $H^2$,
\be\label{PSEC} < \Psi_{X h} , \Psi _{Yh}> =e^{-{1\over 4h}(|X-Y|^2 ) + {i\over 2h} \sigma (X , Y) }.
\ee
where  $\sigma$ is the symplectic form $\sigma ( ( q , p), (q', p')) = p \cdot q' - q \cdot p'$. 
The following  proposition is a consequence of this definition together with the fact that
$S_{h, 2 }$ is a unitary  isomorphism.

 \begin{prop}\label{isom}  i)  The operator  $ T^{FF}$ is a partial isometry from
 ${\cal F}_s (H_{\bf C})$ into  ${\cal F}_s (H_{\bf C} \times H_{\bf C} )$.
For all $R>1$ and for any $f$ in   ${\cal F}_s^{R} (H_{\bf C})$, we have
with the  notation  (\ref{norme-FsR}):
 $$  \Vert T^{FF}f \Vert_{R, 2} = \Vert f \Vert_{R} $$
 ii) The map $T_h^{FW}$ is a partial isometry from   ${\cal F}_s (H_{\bf C})$ into
  $L^2(B^2 , \mu_{B^2, h})$.

 \end{prop}

The next  Proposition is used  Section \ref{s2-3}.

 \begin{prop}\label{prop-SB}  For any $V $ in  $H$ being identified with
the element $(V, 0)$ and for all $X= (q , p)$ in  $H^2$, we have:
 \be\label{crea-FH} T^{FH} ( a^{\star}( V) f)  (q , p) = \frac{1}{ \sqrt {2h}}   ( V \cdot q - i V \cdot p) (T^{FH}f) (q, p), \ee
 \be\label{ann-FH} T^{FH} ( a( V) f)  (q , p) = \sqrt { \frac {h} {2}} ( V \cdot ( \partial _q + i
 \partial _p)) (T^{FH}  f)  (q , p). \ee

 \end{prop}

 {\it Proof.}   Let us set $ (q+ip)^{\otimes m} = (q+ip)\otimes \cdots \otimes (q+ip)$
 ($m$ times). Using the formulas (X.63) of \cite{RSII}, we have:
 $$ a(V) (q+ip) ^{\otimes m}  = \sqrt {m}\ ( V\cdot (q+ip)) \ (q+ip)) ^{\otimes m-1}, $$
 $$ a^{\star} (V) (q+ip) ^{\otimes m}  = \frac {1} {\sqrt {m+1} } \sum _{k=0}^m
 (q+ip ) ^{\otimes k} \otimes V \otimes (q+ip ) ^{\otimes m- k}. $$
 In other words:
 $$ V \cdot ( \partial _q - i \partial _p)(q+ip) ^{\otimes m}  =
 2 \sqrt {m} a^{\star} (V) (q+ip) ^{\otimes m-1}. $$
 By (\ref{C-S}),  it follows that:
$$ a(V) \Psi_{q , p, h} =  \frac{1}{ \sqrt {2h}}   V\cdot (q+ip) \Psi_{q , p, h} $$
 and also:
 $$ a^{\star}(V) \Psi_{q , p, h} = \sqrt { \frac {h} {2}} e^{-(|q|^2+ |p|^2)  / 4h}
 V \cdot ( \partial _q - i \partial _p) \left (   e^{(|q|^2+ |p|^2) / 4h}  \Psi_{q , p, h} \right ).  $$

Equalities (\ref{crea-FH}) and (\ref{ann-FH}) follow easily by (\ref{TFH}).

   \fpr

   \subsection{Power series expansions. }\label{s2-3}

We give here the power series expansions of the two functions $T_h ^{FH}f$ and $T_h ^{FW}f$
 with two different convergence senses and  for any $f$ in  ${\cal F}_s (H_{\bf C})$.

Fix a Hilbertian basis $(e_j)$ of  $H$. Then define the basis
$(u_{\alpha })$ of   ${\cal F}_s (H_{\bf C})$ as in (\ref{u-alpha}).
For each multi-index
 $\alpha$, also define a  function $ \Phi_{\alpha , h}$ on  $H^2$ by:
 \be\label{Phi-alpha}  \Phi_{\alpha ,  h } (q , p) = (2h) ^{-|\alpha|/2} (\alpha ! )^{-1/2}
   \prod _j ( e_j\cdot (q - ip)) ^{\alpha _j}.\ee

The convergence concerning the expansion of the function $T_h ^{FH}f$ is pointwise and not absolute due to the infinite dimensional setting.
Let ${\cal M} $ be the set of all multi-indices $\alpha$ and let  $(I_N)$ be an increasing sequence of finite subsets of  $I$ with union  ${\cal M}$.

 \begin{prop}\label{DSE} i) For any $f$ in  ${\cal F}_s (H_{\bf C})$ and for all $X$ in
 $H^2$, we have:
\be\label{DSFH} ( T_h^{FH}f)  (X) =  \lim _{N \rightarrow \infty}  \sum _{\alpha  \in I_N} < f ,  u_{\alpha } >
 \Phi_{\alpha , h} (X).    \ee
 ii) The  function $T_h^{FH}f$ is Gateaux anti-holomorphic if $H^2$ is identified with $H_{\bf C}$ (when identifying
 $X=(q, p)$ with $q+ip$). That is, for any complex subspace  of  $H_{\bf C}$ of
  finite dimension, the  restriction of  $T_h^{FH}f$ to $E$ is anti-holomorphic.

 iii) For any square-summable sequence $(a_{\alpha})$, there is a  function $F$ on
 $H^2$ satisfying  for all $X$ in  $H^2$
 \be\label{conv-simple1}  F(X) =
   \lim _{N \rightarrow \infty}  \sum _{\alpha \in I_N  } a_{\alpha } \Phi_{\alpha , h} (X). \ee
Moreover, there exists $f$ in  ${\cal F}_s (H_{\bf C})$ satisfying  $T_h^{FH} f = F$.

 \end{prop}

 {\it Proof.} Let us first prove that :
  \be\label{ualphaPhialpha} T_h^{FH} u_{\alpha } (X) =   e^{ \frac {|X|^2 } {4h}} < u_{\alpha } , \Psi _{X , h} > = \Phi_{\alpha , h} (X)
  \hskip 2cm X = (q , p) \in H^2. \ee
This point is proved by induction on   $|\alpha|$.
For $\alpha = 0$, namely  for $u_{0}= \Psi_0$ and $\Phi_{0, h} = 1$,
it is a consequence of  (\ref{TFH}). It is then obtained by iteration using (\ref{crea-FH}).

Concerning point i), set for any integers $N$:
\be\label{f-N} f_N = \sum _{\alpha \in I_N  } < f ,  u_{\alpha } >   u_{\alpha }.\ee
According to (\ref{ualphaPhialpha}), one has for any $X$ in  $H^2$:
 $$  \sum _{\alpha  \in I_N  } < f ,  u_{\alpha } >
 \Phi_{\alpha , h} (X) =  e^{ \frac {|X|^2 } {4h}} < f_N , \Psi _{X , h} > .  $$
Since $f_N$ converges to $f$ in  ${\cal F}_s (H_{\bf C})$ then  point i) follows,
 for any fixed $X$.

 ii) Let $E$ be a complex subspace  of  $H_{\bf C}$ of  finite dimension $m$. Set
 $(w_1 , ... , w_m)$ a complex orthonormal basis of  $E$. Thus, any element
 $Z$ of  $E$ is written as  $Z = z_1 w_1 + \cdots+z_m w_m$, with  $z_j$ in
 ${\bf C}$. Thanks to (\ref{Phi-alpha}) and (\ref{DSFH}), we can then write, for all $Z$ in  $E$:
 $$ T_h^{FH}f (Z) = \sum c_{\gamma } \overline z ^{\gamma}$$
 where  the  series in the right hand side is converging   for all $z$ in  ${\bf C}^m$. The  anti-holomorphic property of  the  restriction of  $ T_h^{FH}f$ to $E$
then follows.

 iii) If the sequence $(a_{\alpha })$ is square-summable, let $f_N$ be the element of
${\cal F}_s (H_{\bf C})$ defined  by:
$$ f_N = \sum _{\alpha \in I_N} a_{\alpha }   u_{\alpha }.$$
The sequence $(f_N)$ tends to an element $f$ of  ${\cal F}_s (H_{\bf C})$ and the above reasoning
shows that:
 $$\lim _{N\rightarrow \infty}\sum _{\alpha \in I_N  } a_{\alpha } \Phi_{\alpha , h} (X)
 = T_h^{FH}f (X).$$
\fpr

Let us now  consider the series expansion of  $T_h^{FW}f$.
One defines a function  $\widetilde  \Phi _{\alpha }$ almost everywhere
on  $B^2$ by:
\be\label{Phi_alpha}\widetilde  \Phi _{\alpha , h } (q , p) = (2h) ^{-|\alpha | /2} (\alpha ! )^{-1/2}
\prod ( \ell _{e_j} (q)  - i \ell _{e_j} (p) )^{\alpha _j}.\ee
The functions $\widetilde  \Phi _{\alpha , h }$ constitute an orthonormal system in
$L^2 ( B^2 , \mu _{B^2, h})$.

\begin{prop} One has:
  \be\label{ualpha-tilde-Phialpha} T_h^{FW} u_{\alpha }  =  \widetilde  \Phi_{\alpha , h}.
  \ee
\end{prop}

{\it Proof.}  If  $(e_j)$ is a Hilbertian basis of  $H$ then we use the   Hilbertian basis
 of  $H^2$ constituted of  the $(e_j, 0)$ and $(0, e_j)$. The elements of
 ${\cal F}_s(H_{\bf C} \times H_{\bf C})$ constructed with  this basis defined as in
   (\ref{u-alpha}) are denoted by $u_{\alpha \beta}$. We observe that:
   $$  \Gamma (U_-) u_{\alpha} =
   2 ^{-|\alpha |/2}  \sum _{\beta + \gamma = \alpha}
\sqrt { \frac {  \alpha !  }  { \beta ! \gamma ! } } (-i )^{|\gamma |} u_{\beta, \gamma }.$$
 According to the Segal isomorphism definition in (\ref{def-segal}), considering $(q, p)$ as the  variable of  $B^2$
 and  setting $\ell _{(e_j , 0)} (q , p) = \ell _{e_j } (q)$
 and $\ell _{(0, e_j )} (q , p) = \ell _{e_j } (p)$, one checks:
 $$ S_{h, 2}  ( u_{\beta , \gamma} )  (q, p) = \prod _{j, \beta _j + \gamma_j \not = 0}
 H_{\beta _j} \Big ( \ell _{e_j } (q) / \sqrt {h} \Big )
 \ H_{\gamma _j} \Big (\ell _{e_j } (p) / \sqrt {h} \Big ) $$
 where  the $H_k$ are  the Hermite polynomials chosen being an orthonormal system in  $L^2(\R , \mu _{\R, 1})$.
 Then we use the following identity, probably standard, for all
 $(x , y)$ in $\R^2$:
 $$ \frac {(x - iy )^m} {\sqrt {m!}}  = \sum _{p+q = m}
 \sqrt {\frac {m! } { p! q! }}
 (-i)^{q} H_q(x) H_{q} (y).$$
The proof of the Proposition is then completed.  \fpr

The result below will be used in   Section \ref{s3}.

  \begin{prop}\label{DSFW} For any $f$ and $g$ in  ${\cal F}_s (H_{\bf C})$,
  one has in the sense of the   $L^2 ( B^2 , \mu _{B^2 , h})$ Hilbert space convergence:  
  \be\label{serie FW}  T_h^{FW}f = \lim _{N \rightarrow \infty}  \sum _{\alpha  \in I_N}
  < f ,  u_{\alpha } > \widetilde  \Phi_{\alpha , h}    \ee
  \be\label{serie FW-2} <f ,g > = \sum < T_h ^{FW}f,  \widetilde \Phi_{\alpha , h} >_h   \
  < \widetilde \Phi_{\alpha , h} ,      T_h^{FW}g>_h   \ee
  where the series is absolutely converging.

   \end{prop}
The above scalar product is the canonical $L^2 ( B^2 , \mu _{B^2 , h})$ scalar product.

Again, we emphasize that the convergence in  (\ref{serie FW}) is the  $L^2 ( B^2 , \mu _{B^2 , h})$ Hilbert space convergence whereas it is a pointwise convergence  in  (\ref{DSFH}), for all $X$ in  $H^2$.

   {\it Proof.}  Let  $f_N$ be the element defined  in (\ref{f-N}).
 One notices in view of (\ref{ualpha-tilde-Phialpha}) that:
   $$  T_h^{FW} f_N =  \sum _{\alpha  \in I_N}
  < f ,  u_{\alpha } > \widetilde  \Phi_{\alpha , h} .   $$
According to point $ii)$ in  Proposition \ref{isom}, one sees that $  T_h^{FW} f_N$  tends to  $  T_h^{FW} f$ in  $L^2 ( B^2 , \mu _{B^2 , h})$. One then gets equality
(\ref{serie FW}).  The second equality therefore holds true since the
 set of functions $\widetilde  \Phi _{\alpha , h }$ is an orthonormal system in
$L^2 ( B^2 , \mu _{B^2, h})$.

\subsection{Connection between  $T_h^{FH}f$ and $T_h^{FW}f$. }\label{s2-4}

In the finite dimensional framework, the   Bargmann transform of a function $f\in L^2(\R^n)$
 is a function belonging to  $L^2(\R^{2n})$ (with suitable measures)  holomorphic
 when $\R^{2n}$ is identified with ${\bf C}^n$. Here, in infinite dimension,  the  function $T_h^{FH}f$
is Gateaux anti-holomorphic when $(q , p)$ is identified with $q+ip$
 (Proposition \ref{DSE}) and $T_h^{FW}f$ belongs to  an $L^2$ space on a suitable  $B^2$  for a convenient mesure.
One then should specify the relation between these two functions. We underline that one cannot consider the restriction to  $H^2$ of a  measurable function defined on  $B^2$ since
 the measure of $H^2$  in  $B^2$ is zero.

We prove that  $T^{FW}_h f$ is a stochastic extension of  $T^{FH}_h f$ in the sense of the definition below.

 With each finite dimensional space  $E\subset H$, one can associate a map
$ \widetilde \pi _E: B \rightarrow  E$ defined almost everywhere by:
 $$ \widetilde \pi _E (x) = \sum _{j=1} ^{{\rm dim} (E ) }
    \ell _{u_j}  (x)u_j  $$
where the $u_j$ ($1 \leq  j \leq  {\rm dim}(E))$  form an orthonormal basis
of $E$ and $\ell _{u_j}$  is defined in Section \ref{s2-1}. This
map does not depend on the choice of the orthonormal basis.

\begin{defi}\label{def-ext-stoc} Let $H$ be a separable real   Hilbert space of infinite dimension.
Set $B$ a Wiener extension of   $H$ and set
$\mu _{B, h}$ the  measure defined in Section \ref{s2-1}.
One says that a measurable function $f$  on  $H$ admits a stochastic extension
in  $L^p (B , \mu _{B , h})$ ($1\leq p <\infty$)  if the  sequence of  functions
 $F \circ \widetilde \pi _{E_n} :  B \rightarrow {\bf C}$ is a Cauchy sequence in
$L^p( B  , \mu _{B , h})$ for any increasing sequence of finite dimensional complex subspaces   $(E_n)$ of  $H$ having a dense union in  $H$. The limit, which is independent of  the  sequence $(E_n)$,
is called the stochastic extension of  $F$ and is here often denoted by $\widetilde F$.
\end{defi}

We also prove in the following that  $T_h^{FH}f$ is the image of $T_h^{FW}f$ using  the reproducing kernel defined below.

For all $X = (q, p)$ in  $H^2$, one defines a
function $\ell_{X} $ almost everywhere on  $B^2$ by   $\ell_{X }(Y) =\ell_{q - i p}(y)+i\ell_{q-ip }(\eta)$
with $Y = (y, \eta)$.
One also defines almost everywhere a function $Y=(y, \eta) \rightarrow B_h(X,Y)$ by:
$$
B_h(X,Y)= e^{\frac{1}{2h}\ell_{X}(Y) }.
$$
One knows that  the  function $Y \rightarrow B_h(X,Y)$
belongs to  $L^2(B^2 , \mu_{B^2, h})$ for any  $X$ in  $H^2$ in such a way that, for all
$F$ in  $L^2(B^2 , \mu_{B^2, h})$,  the following integral called reproducing kernel:
$$ (B_hF)  (X) =
\int _{B^2}  B_h (X , Y)   F (Y) d\mu _{B^2,h} (Y)
$$
defines a function $ B_h F$  on  $H^2$.

The link between $T^{FH}_h f$ and $T^{FW}_h f$ is given by the  proposition below.

  \begin{prop}\label{2-transfo} i)
  There exists a stochastic extension in    $L^2(B^2 , \mu_{B^2, h})$ of the
 function   $T_h^{FH}f$ defined on  $H^2$, for all $f\in {\cal F}_s (H_{\bf C})$. This stochastic extension is  $T_h^{FW}f$.

 ii) One has,
\be\label{reproTh} (T_h^{FH}f) (X) = \int _{B^2}  B_h (X , Y)   T_h^{FW}f (Y) d\mu _{B^2,h} (Y)
\ee
for every $f$ in  ${\cal F}_s (H_{\bf C})$ and for all
$X$ in  $H^2$.

 \end{prop}

The proof of point i) relies on the next theorem, which is a variant of Theorem 8.8 in \cite{A-J-N-1} adapted to our current  situation.

 \begin{theo}\label{th-ext-stoc} Let $H$ be an infinite dimensional separable complex Hilbert space and set
 $B$ a Wiener extension of  $H$. The space $B$ is considered as a real space. Fix a continuous function $F$  defined on  $H$. We assume that:

 i) The function $F$ restricted to
 $E$ is anti-homomorphic where $E$ is any finite dimensional complex subspace of $H$.

 ii) For every finite dimensional complex subspace  $E$ of $H$,  the  function $F$ restricted to
 $E$ belongs to  $L^2 (E , \mu _{E , h} )$ and its norm is bounded independently of  $E$
 ($\mu _{E , h}$ denotes the Gaussian measure with  variance $h$  on  $E$  considered as a real space).

Then, there exists $\widetilde F$ a stochastic extension in  $L^2(B , \mu _{ B , h})$ of the function $F$ (in the sense of
Definition \ref{def-ext-stoc}).
Moreover, we have:
\be\label{norme-ext} \Vert \widetilde F \Vert _{ L^2 (B  , \mu _{B , h})} \leq
\sup _E       \Vert R_E F \Vert _{ L^2 (E , \mu _{E , h})} \ee
where  $R_E  F$ stands for  $F$ restricted to $E$ and where  the supremum is running over
all the finite dimensional complex subspaces $E$  in  $H$.

 \end{theo}

 {\it Proof of Proposition \ref{2-transfo}.}  i) For all $f$ in  ${\cal F}_s (H_{\bf C})$,
 we know (Proposition \ref{DSE})  that  the  function   $ T_h^{FH}f$, defined on  $H^2$ is identified to
 a function $F$ being Gateaux anti-holomorphic  on  $ H_{\bf C}$ when $H^2 $
 and $H_{\bf C}$ are identified. One also knows that  this  function satisfies hypothesis ii)
 of Theorem \ref{th-ext-stoc} and, for all complex subspaces $E$ of
  $H_{\bf C}$:
 $$  \Vert R_E F \Vert _{ L^2 (E , \mu _{E , h})}  \leq \Vert f \Vert. $$
 This fact comes from the standard properties
 of  the Segal Bargmann transform in finite dimension. According to Theorem \ref{th-ext-stoc}, $T_h^{FH}f$ has a stochastic extension
 $\widetilde F$ in  $ L^2 (B^2 , \mu _{B^2 , h})$. Let $(f_N)$ be a sequence in   ${\cal F}_s (H_{\bf C})$
 converging to $f$ where each $f_N$ is a linear combination of the
 $u_{\alpha}$. From (\ref{ualphaPhialpha}) and (\ref{ualpha-tilde-Phialpha}),
 $T_h^{FW} f_N$ is the stochastic extension of  $T_h^{FH} f_N$.  The  sequence
  $T_h^{FW} f_N$ tends to $T_h^{FW} f$ in   $ L^2 (B^2 , \mu _{B^2 , h})$. For all
 complex subspace $E$ of  finite dimension in  $H^2$, one has:
  $$  \Vert R_E  T_h^{FH} (f_N -f)  \Vert _{ L^2 (E , \mu _{E , h})}  \leq \Vert f_N -f \Vert.$$
Consequently, in view of Theorem \ref{th-ext-stoc}, the  sequence of the
  $T_h^{FW} f_N$,  stochastic extensions of the $T_h^{FH} f_N$, converges to the
  stochastic extension of  $T_h^{FH} f$, which is therefore equal to $T_h^{FW} f$.

 ii) With the above notations, the Segal Bargmann standard properties
within the finite dimensional framework show that, for all $X$ in  $H^2$ and for
any $N$:
$$  T_h^{FH}f_N (X) = \int _{B^2}  B_h (X , Y)   T_h^{FW}f_N (Y) d\mu _{B^2,h} (Y).
$$
From Definition (\ref{TFH}), for all $X$ in  $H^2$, the  sequence
 $T_h^{FH}f_N (X)$ converges to $T_h^{FH}f (X)$. Besides, the  sequence
 $(T_h^{FW}f_N) $ tend vers   $T_h^{FW}f $ in   $ L^2 (B^2 , \mu _{B^2 , h})$ and we consequently have, since the  function $Y\rightarrow  B_h (X , Y)$ belongs to  $ L^2 (B^2 , \mu _{B^2 , h})$:
$$ \lim _{N\rightarrow \infty}
\int _{B^2}  B_h (X , Y)  T_h^{FW}(f_N - f) (Y) d\mu _{B^2,h} (Y)= 0.
$$
One then obtains (\ref{reproTh}).   \fpr

\subsection{Segal Bargmann transform and the product law $I$. }\label{s2-5}

We first start with:

{\it Proof of  Proposition \ref{norme-prod-C}.}
We consider the  Hilbertian basis $(u_{\alpha})$ of  (\ref{u-alpha}),
  and equality (\ref{C-basis-2}).
Setting $\varphi = I(f , g)$, we have, for all multi-indices $\gamma$:
$$ < \varphi , u_{\gamma} > = \sum _{\alpha + \beta = \gamma }
 < f, u_{\alpha} >   < g , u_{\beta} > \left [ \frac { \gamma !}
{\alpha ! \beta! } \right ]^{1/2}. $$
From Cauchy-Schwarz:
$$ (R'') ^{|\gamma |} | < \varphi , u_{\gamma} > |^2 \leq
\sum _{\alpha + \beta = \gamma }  R^{\alpha }  | < f , u_{\alpha} > |^2
 (R')^{\beta }  | < g , u_{\beta} > |^2 \
 \sum _{\alpha + \beta = \gamma }  \frac { \gamma !}
{\alpha ! \beta! } \ \frac {  (R'') ^{|\gamma |}  }
{   R ^{|\alpha |}      (R') ^{|\beta |}  }.$$
If  $(1/R'') = (1/R) + (1/R')$ then the latter sum equals to $1$.
One verifies that $f$  belongs to ${\cal F}_s^{R} (H_{\bf C})$  if and only if
\be\label{norme-DR}  \Vert f \Vert _R ^2 = \sum |< f ,  u_{\alpha } > |^2 R^{|\alpha |} < \infty. \ee
Therefore, the  Proposition is proved.  \fpr

 \begin{theo}\label{prod-SB}  i) For all $h>0$, for any $f$ in  ${\cal F}_s^{R} (H_{\bf C})$
  and $g$ in  ${\cal F}_s^{R'} (H_{\bf C})$, ($R\geq 1$, $R'\geq 1$,
  $(1/R) + (1/R') \leq 1$), one has:
  $$ T_h ^{FH} I(f , g) =  (T_h ^{FH} f) \ (T_h ^{FH} g) $$
  where the composition in the right hand side refers to the ordinary product  of functions defined on  $H^2$.

  ii) One also has, for any Wiener extension $B$ of  $H$:
  $$ T_h ^{FW} I(f , g)(X) =  (T_h ^{FW} f)(X) \ (T_h ^{FW} g)(X)
  \hskip 2cm a.e. X\in B^2. $$
 \end{theo}

 {\it Proof. Point i)}    We use  the  Hilbertian basis
 $(u_{\alpha})$ of  (\ref{u-alpha}).  From (\ref{u-alpha}) and (\ref{equ-prod}),
one sees:
 \be\label{C-basis-2} I( u_{\alpha} ,  u_{\beta} ) = \left [ \frac { (\alpha + \beta )!}
{\alpha ! \beta! } \right ]^{1/2}     u_{\alpha + \beta}. \ee
Therefore, if  $f$ and $g$ belong to  ${\cal F}_s^{\rm fin} (H_{\bf C})$ (defined in Section \ref{s1}),
$$ T_h^{FH} I(f , g) (X) = \sum _{\alpha , \beta }
 < f , u_{\alpha} >\  < g , u_{\beta} > \
\left [ \frac { (\alpha + \beta )!}
{\alpha ! \beta! } \right ]^{1/2}      T_h^{FH}  u_{\alpha + \beta} (X). $$
According to  Proposition \ref{DSE}:
$$  T_h^{FH} f (X) = \sum _{\alpha }  < f , u_{\alpha} >
\Phi_{\alpha , h}   (X).$$
From (\ref{Phi-alpha}), one verifies:
$$ \sqrt { (\alpha + \beta) !} \Phi_{\alpha + \beta  , h} (X) =
 \sqrt { \alpha !}  \Phi_{\alpha , h} (X) \  \sqrt { \beta !}  \Phi_{\beta , h} (X).$$
{\it Point ii)}
 One knows by Proposition \ref{2-transfo} that
$T_h ^{FH} I(f , g) , T_h ^{FH} f $ and $ T_h ^{FH} g$
have the stochastic extensions
$T_h ^{FW} I(f , g) , T_h ^{FW} f $ and $ T_h ^{FW} g$
in the $L^2(B^2, d\mu_{B^2, h})$ sense. Let $(E_n)$ be an increasing
sequence of finite dimensional subspaces of $H$, with a dense union.
Since $T_h ^{FH} f \circ \tilde{\pi}_{E_n}$ and
$T_h ^{FH} g \circ \tilde{\pi}_{E_n}$  converge to $ T_h ^{FW} f $ and
$ T_h ^{FW} g$ in  $L^2(B^2, d\mu_{B^2, h})$, the product
$T_h ^{FH} f T_h ^{FH} g \circ \tilde{\pi}_{E_n}$
converges to
$ T_h ^{FW} f  T_h ^{FW} g$ in  $L^1(B^2, d\mu_{B^2, h})$.
The convergence is almost sure for a subsequence indexed by $\varphi(n)$.
By point i), $(T_h ^{FH} f T_h ^{FH} g )\circ \tilde{\pi}_{E_{\varphi(n)}}
  =T_h ^{FH} I(f , g) \circ \tilde{\pi}_{E_{\varphi(n)}} $. Moreover,  it
    converges to $T_h ^{FW} I(f , g) $ in $L^2$.
    Extracting a further subsequence
    gives an almost sure convergence and the equality
    $T_h ^{FW} I(f , g) =(T_h ^{FW} f)(  T_h ^{FW} g) \ a.e.$

 \fpr

{\it Proof of Theorem \ref{t-J}. } We prove equality  (\ref{J-AW}).   We use Definition (\ref{OPAW}).
We have, for all $f$ in  ${\cal F}_s^{R} (H_{\bf C})$  ($R>1$) and for all $g$
in  ${\cal F}_s (H_{\bf C})$, according to Proposition  \ref{isom} and Definition
(\ref{def-J-lambda}):
$$ < J_{\lambda }  f , g> = \int _{B^2} \big ( T^{FW}_1  I( U_{\lambda } ,  f) \big ) (x)
\ \overline { (T^{FW}_1 g) (x)} d \mu _{B^2 , 1} (x).  $$
In view of Theorem \ref{prod-SB}:
$$ < J_{\lambda }  f , g> = \int _{B^2}  (T^{FW}_1 U_{\lambda }) (x) \
(T^{FW}_1 f) (x) \ \overline { (T^{FW}_1 g) (x)} d \mu _{B^2 , 1} (x).  $$
From Theorem \ref{2-transfo}, the  function $T^{FW}_1U_{\lambda }$
 is indeed the stochastic extension of  $F_{\lambda } = T^{FH}_1 U_{\lambda }$,
 which formally proves equality (\ref{OPAW}). The  convergence
 of  the integral is handled by the fact that, for all $f$ in  ${\cal F}_s^{R} (H_{\bf C})$  ($R>1$),
if  $R'$ satisfies $(1/R) + (1/R') = 1$ then we know that   $U_{\lambda }$ is in
${\cal F}_s^{R'} (H_{\bf C})$. From  Proposition \ref{norme-prod-C},
$I( U_{\lambda } ,  f)$ is well defined as an element in  ${\cal F}_s (H_{\bf C})$.
Following Proposition \ref{isom},   $T^{FW}_1 I( U_{\lambda } ,  f)$
is well defined as an element of  $L^2 ( B^2 , \mu _{B^2 , 1})$, as
 $T^{FW}_1 g$, ensuring the  convergence of  the integral. \fpr

{\it Proof of equality (\ref{equ-I-EC}). }      We can write, if  $X = (q, p)$:
 $$ \Psi_{X, h }  = \sum _{m\geq 0}
 {e^{-{|X|^2  \over 4h }} \over (2h)^{m/2} m! } \Big ( a^{\star} ( q+ ip)\Big )^m \Psi_0.$$
One checks that the coherent states belong to ${\cal F}_s^R({ H}_{\bf C} )$
for every $R\geq 1$. Let
$ \Psi_{X, h }^N $ be a truncated coherent state, the sum running
on $\{ m\leq N\}$ instead of $\N$.

Proposition \ref{norme-prod-C} with  $R=R'=2$ shows that
$I(\Psi_{X, h }^N,\Psi_{Y, h }^N)$ converges to
$I(\Psi_{X, h },\Psi_{Y, h })$ in  ${\cal F}_s^1({ H}_{\bf C} )=
{\cal F}_s({ H}_{\bf C} )$. The computation below shows that
$I(\Psi_{X, h }^N,\Psi_{Y, h }^N)= \Psi_{X+Y, h }^N + V_N$ and one easily proves that the rest $V_N$ converges to $0$ when $N$ converges to infinity.

 $$ I ( \Psi_{X, h }^N , \Psi_{Y, h }^N) =  \sum _{0\leq m, n \leq N}
  \frac{e^{-\frac{|X|^2 + |Y|^2 }{ 4h }}}{ (2h)^{m+n/2} m! n!}
  I(( a^{\star} ( q+ ip))^m \Psi_0 , ( a^{\star} ( q'+ ip') )^n \Psi_0 ) $$
  $$ =  \sum _{0\leq m, n \leq N}
  \frac{e^{-{|X|^2 + |Y|^2  \over 4h }} }{(2h)^{m+n/2} m! n!}
 ( a^{\star} ( q+ ip) )^m  ( a^{\star} ( q'+ ip'))^n  \Psi_0  $$
  $$
  = e^{-\frac{|X|^2 + |Y|^2 }{ 4h }}  \sum _{s =0}^N
  \frac{1}{s! (2h)^{s/2}} \sum_{m=0}^s \frac{s!}{m! (s-m)!}
   ( a^{\star} ( q+ ip) )^m  ( a^{\star} ( q'+ ip'))^{s-m}  \Psi_0
  + ...
  $$
  $$
 ... +
   e^{-\frac{|X|^2 + |Y|^2 }{ 4h }}  \sum _{s =N+1}^{2N}
  \frac{1}{s!  (2h)^{s/2}} \sum_{m=s-N}^{N} \frac{s!}{m! (s-m)!}
   ( a^{\star} ( q+ ip) )^m  ( a^{\star} ( q'+ ip'))^{s-m}  \Psi_0
  $$
   $$
  = e^{-\frac{|X|^2 + |Y|^2 }{ 4h }}  \sum _{s =0}^N
  \frac{1}{s! (2h)^{s/2}}
   ( a^{\star} ( q+ q'+ i(p+ p')) )^s   \Psi_0
 \ +\  V_N = \Psi_{X+Y, h }^N + V_N.
  $$
One then deduces (\ref{equ-I-EC}). \fpr

 \section{Absolute convergence of the  Mizrahi series.}\label{s3}

If $A$ is a bounded operator on ${\cal F}_s (H_{\bf C})$ or an unbounded
operator $(A , D(A))$ such that the domain $D(A)$ contains the coherent states,
then the Wick symbol of $A$ is the  the function  $\sigma_h^{wick} (A)$ defined on
$H^2$ by:
\be\label{1.3}\sigma_h^{wick} (A) (X) = < A \Psi_{X , h} ,    \Psi_{X , h}>,
\ee
where the coherent states $\Psi_{X , h}$ are defined in (\ref{C-S}).

We shall state and prove  a formula giving the Wick symbol of the
composed $A\circ B$ of two bounded operators  $A$ and $B$ in
${\cal F}_s(H_{\bf C})$, as the sum of a convergent series.
In finite dimension, it is a result of  Mizrahi \cite{Mizrahi}
and Appleby \cite{Appleby}.

If $H$ is a separable, real Hilbert space, if $(e_j)$ is a Hilbert basis of $H$,
we define, for all multi-index  $\alpha = ( \alpha _j) $
(which means $\alpha _j = 0$ except for a finite number of values of $j$),
two differential operators  on $H^2$, denoting by
$(q, p)$ the variable of $H^2$:
$$ (\partial _q \pm i \partial _{p})^{\alpha }
=  \prod _j  \left ( \frac {\partial} {\partial q_j} \pm i
\frac {\partial} {\partial p_j}\right )^{\alpha _j} . $$

\begin{theo}\label{t-comp-Wick}  For every bounded operators  $A$ and $B$ in
${\cal F}_s (H_{\bf C})$,
the Wick symbols $F$ and $G$ of $A$ and $B$ are $C^{\infty }$ functions on $H^2$.  For each $k\geq 0$, the following series, a priori defined using a Hilbertian basis $(e_j)$  of $H$:
\be\label{CNWick}
C_k^{wick} (F, G) = 2^{-k} \sum_{|\alpha | = k}  \left (\frac  {1}{  \alpha !} \right )
(\partial _q - i \partial _{p})^{\alpha }  F  \
( \partial _q + i \partial _{p })^{\alpha } G
\ee
is absolutely convergent and its sum $C_k^{wick} (F, G)$ is independent of the basis.
We have, for all $X$ in $H^2$:
\be\label{aaa} \sigma_h^{wick} (A \circ B) (X) = \sum _{k= 0}^{\infty}
h^k C_k^{wick} (  \sigma_h^{wick} (A) , \sigma_h^{wick} (B) ) (X),
\ee
where the series is absolutely convergent.

\end{theo}

See \cite{Appleby} or \cite{Mizrahi} in the case of finite dimension.
See  also \cite{A-N},   formula (15)(i).
If the operators $A$ and $B$ act in  ${\cal F}_s (H_{\bf C})\otimes E$
where  $E$ is  a  Hilbert of  finite dimension then the functions $F$ and $G$
are defined on  $H^2$, taking values in  ${\cal L}(E)$ and the product in the right hand side
of  (\ref{aaa}) is a composition of operators.
   Note that, if one of the functions $F$ or $G$ is scalar valued, one has
$$ C_1^{wick} ( F , G) - C_1^{wick} ( G , F) = i^{-1} \{ F , G \} .$$
One denotes by  $\sigma$ the symplectic form on $H^2$ defined by
$\sigma ( (x , \xi) ,
(y , \eta)) = y \cdot \xi - x \cdot \eta $ and by  $\{ F , G \}$
the Poisson bracket of two functions $F$ and $G$ in
$C^1(H^2)$, defined by $\{ F , G \} (X)
= - \sigma ( dF(X), dG(X)) $.

{\it Proof of Theorem \ref{t-comp-Wick}.}
It suffices to prove the following equality, for all $X$ in $H^2$:
\be\label{2.2}  <  B \Psi_{Xh} , A^{\star}  \Psi_{Xh} > = \sum
 \frac { h  ^{|\alpha |} } { 2^{|\alpha |} \alpha ! } \
 (\partial _q - i \partial _{p})^{\alpha }  \sigma_h^{wick} (A) (X)\
 (\partial _q + i \partial _{p})^{\alpha }  \sigma_h^{wick} (B) (X)
\ee

{\it Proof of  (\ref{2.2})  for $X=0$.}
By Proposition \ref{DSFW}, we have:
$$  <  B \Psi_{0h} , A^{\star}  \Psi_{0h} > = \sum
< T_h ^{FW}B \Psi_{0h}  , \widetilde \Phi_{\alpha , h}>
< \widetilde \Phi_{\alpha , h} , T_h ^{FW}  A^{\star}  \Psi_{0h}  >.  $$
Applying Proposition \ref{2-transfo} to $B \Psi_{0h}$, we obtain, for $X$ in $H^2$:
$$ T_h^{FH}(B \Psi_{0h}) (X)  = \int _{B^2}  {\cal B}_h (X , Y) \
T_h^{FW}(B \Psi_{0h}) (Y) \   d\mu _{B^2, h} (Y).$$
Differentiating the integral,
$$ ( \partial _{x_j} + i \partial _{\xi_j})  T_h^{FH}(B \Psi_{0h}) (X) = h^{-1}  \int _{B^2}
\ell _{e_j } ( y+ i \eta){\cal B}_h (X , Y)\
T_h^{FW}(B \Psi_{0h}) (Y)
\  d\mu _{B^2, h} (Y). $$
Iterating and then setting $X= 0$, one obtains
$$ ( \partial _x + i \partial _{\xi })^{\alpha }  T_h^{FH}(B \Psi_{0h}) (0) =
h^{-|\alpha| }  \int _{B^2}
\left [\prod ( \ell _{e_j} (y)  + i \ell _{e_j} (\eta) )^{\alpha _j} \right ]\
T_h^{FW}(B \Psi_{0h}) (Y)
\ d\mu _{B^2, h} (Y). $$
Equivalently,
$$ < T_h^{FW}(B \Psi_{0h}) , \widetilde \Phi_{\alpha , h} > =\left ( {h\over 2} \right ) ^{|\alpha |/2} (\alpha !)^{-1/2}
( \partial _x + i \partial _{\xi })^{\alpha } T_h^{FH}(B \Psi_{0h}) (0) $$
and the same equality holds for $T_h^{FW}(A^{\star}  \Psi_{0h}) $. Consequently,
$$  <  B \Psi_{0h} ,  A^{\star} \Psi_{0h} > = \sum \left ( {h\over 2} \right ) ^{|\alpha |} (\alpha !)^{-1}
( \partial _x + i \partial _{\xi })^{\alpha }T_h^{FH}(B \Psi_{0h})  (0)  \
\overline {( \partial _x + i \partial _{\xi })^{\alpha }  T_h^{FH}(A^{\star}  \Psi_{0h})(0)}.
$$

For any bounded operator $B$ in ${\cal F}_s (H_{\bf C})$, for all $X$ and $Y$ in $H^2$, set:
\be\label{2.4} (S_hB) (X , Y) = { < B \Psi_{X h}, \Psi_{Yh} > \over <  \Psi_{X h}, \Psi_{Yh} > }.
\ee
By (\ref{PSEC}) and (\ref{TFH}), we have:
$$ (S_hB) (X , Y) = e^{\frac {|X|^2} {4h} } \   e^{- \frac {1} {2h}  (x+ i \xi) \cdot ( y-i \eta) }
 \Big ( T_h ^{FH} ( B \Psi_{X h})  \Big ) (Y)$$
 Therefore, when $Y = (y, \eta) $ is identified with $y+i\eta$, the function
 $Y \rightarrow (S_hB) (X , Y)$ is Gateaux anti-holomorphic. Similarly, the function
 $X \rightarrow (S_hB) (X , Y)$ is Gateaux holomorphic.
 Since the restriction of this function to the diagonal is the Wick
 symbol of $B$, one has,
$$ ( \partial _q + i \partial _{p })^{\alpha }T_h^{FH}(B \Psi_{0h})  (0)  =
( \partial _q + i \partial _{p })^{\alpha }
\sigma_h^{wick} (B) (0).$$
The equality  (\ref{2.2}) is then proved for $X=0$.

{\it Proof of  (\ref{2.2})  for arbitrary $X$.}
The coherent states $\Psi_{X, h}$ defined in (\ref{C-S})  satisfy classically
 \be\label{alter-EC}  \Psi_{X, h} = V_h(X)  \Psi_0
 \hskip 2cm
  V_h(X) = e^{- \frac {i} {\sqrt {h} } \Phi _S ( \widehat  X )}\ee
 where $\Psi_0$ is the vacuum (independent of $h$),  $ \widehat X = (-b, a) $
 for $X = (a, b)$, and $\Phi _S ( \widehat  X )$ is the Segal field associated
 with $ \widehat X  \in H^2$.
We know that, for all $U$ and $V$ in $H^2$:
  \be\label{compos-Segal} e^{i   \Phi _S ( U )}  e^{i   \Phi _S ( V )}
  =  e^{ \frac{i}{2} \sigma ( U ,   V  )  }  e^{i   \Phi _S ( U +V )}\ee
  where  $\sigma $ is the symplectic form $\sigma ( (a, b) , (q, p)) =
  b\cdot q - a \cdot p$.
  Using  (\ref{alter-EC}) and (\ref{compos-Segal}), we obtain:
$$ \sigma_h^{wick} \Big (V_h(X) A V_h(-X) \Big ) (Y) = \Big ( \sigma_h^{wick} (A) \Big ).
(Y-X).$$
One then deduces equality  (\ref{2.2}) for any arbitrary $X$,  and
 Theorem \ref{t-comp-Wick} is proved.

\section{Weyl symbol composition.}\label{s4}

In  \cite{A-L-N-3}, one shows that, if  $F$ and $G$ are in
$S(H^2 , Q)$ (see Definition \ref{d1.1}) and when  $A_Q$ (see (\ref{1.100})) is trace class then the composition $Op_h ^{weyl} (F) \circ  Op_h ^{weyl} (G)$
is written   as $Op_h ^{weyl}(K_h^{weyl} (F, G) )$ with $K_h^{weyl} (F, G) $ in   $S(H^2,4 Q)$.
Our purpose is now  to write $K_h^{weyl} (F, G) $ as an absolutely convergent series.

One defines a differential operator
$\sigma ( \nabla _1, \nabla _2)$ on $H^2 \times H^2$ by:
\be\label{sigma} \sigma ( \nabla _1, \nabla _2) F= \sum _j {\partial^2 F \over \partial y_j \partial \xi_j }
- {\partial^2 F \over \partial x_j \partial \eta_j }
\ee
where $(x, \xi, y, \eta)$ is the variable in  $H^2 \times H^2$. This
operator is defined only for functions $F$ for which the series converges. For
all integers $k$ and all functions $F$ and $G$ such that the series below
makes sense, set:
\be\label{CNWeyl}
C_k^{weyl} (F, G) (X) =   \frac { 1 }{ (2i)^k k!}
\sigma ( \nabla _1, \nabla _2) ^k (F \otimes G) (X , X).
\ee

\begin{theo}\label{t1.6} Let $Q$ be a  nonnegative quadratic form
on $H^2$ where $A_Q$ is trace class.
 Let $F$  and
$G$ be two functions in   $S  (H^2 , Q) $. Then,

i) For each $k\geq 0$, we have:
  \be\label{MajoCkWeyl} \Vert C_k^{weyl} (F, G) \Vert _{4 Q} \leq
  \Vert F \Vert _{Q }
\Vert G \Vert _{Q }  \frac { ({\rm Tr} A_Q)^k } { 2^k k!}.\ee
The series
$$ K_h^{weyl} (F , G) (X)  = \sum _{k= 0}^{\infty} h^k C_k^{weyl} (F, G) $$
is absolutely convergent and defines a function $K_h^{weyl} (F , G)$ in $S(H^2,4 Q)$.

ii) One has:
\be\label{defcomp} Op_h ^{weyl} (F) \circ  Op_h ^{weyl} (G) =  Op_h ^{weyl}
(K_h^{weyl} (F , G)).\ee
iii) For all integers $M$, one can write
\be\label{DASWeyl}
K_h ^{weyl} (F , G) =  \sum _{k= 0}^M h^k   C_k^{weyl} (  F,   G ) + h^{M+1} R_M^{weyl} ( F , G, h)
\ee
with:
$$ \Vert  R_M^{weyl} ( F , G, h) \Vert _{4 Q} \leq  \Vert F \Vert _{Q }
\Vert G \Vert _{Q }   {( {\rm Tr} A_Q)^{M+1} \over (M+1)!} \  e^{(h/2) {\rm Tr} A_Q }.$$
\end{theo}

{\it Proof. i) and iii)} According to (\ref{sigma}), we have, choosing a Hilbertian basis $(e_j)$:
$$ \Vert \sigma ( \nabla _1, \nabla _2) ^k (F \otimes G) \Vert _{\infty}  \leq
\Vert F \Vert _{Q }  \Vert G \Vert _{Q } \left [ \sum _j 2 Q(e_j , 0)^{1/2}
 Q(0, e_j )^{1/2}\right ]^k $$
 $$ \leq \Vert F \Vert _{Q }  \Vert G \Vert _{Q } \left [ \sum _j (Q(e_j , 0)
 + Q(0, e_j ) ) \right ]^k $$
 $$ \leq \Vert F \Vert _{Q }  \Vert G \Vert _{Q } ( {\rm Tr} A_Q)^k. $$
From (\ref{CNWeyl}), one sees:
$$ \Vert  C_k^{weyl} (F, G) \Vert_\infty  \leq    \frac { 1 }{ 2^k k!}
 \Vert \sigma ( \nabla _1, \nabla _2) ^k (F \otimes G) \Vert _{\infty}.  $$
We estimate similarly the derivatives and we deduce (\ref{MajoCkWeyl}).
The series defining  $K_h^{weyl} (F, G) $ in Theorem \ref{t1.6} is then convergent
and the asymptotic expansion iii) is valid.

ii) Equality (\ref{defcomp})
comes from the next Lemma. Indeed, this Lemma shows that
the operators in the left and right hand sides of
 (\ref{defcomp}) share the same  Wick symbol. Thus they are equal.

\bigskip

 \begin{lemm}\label{2-comp} For all $F$ and $G$ in $S(H^2 , Q_A)$, one has:
\be\label{series-form} \sum _{k=0}^{\infty} h^k  C_k^{wick} ( H_{h/2} F , H_{h/2} G )
=
  H_{h/2}\left [  \sum _{k=0}^{\infty}h^k  C_k^{weyl} (F , G) \right ] \ee
where $C_k^{weyl}(F, G)$ is defined in  (\ref{CNWeyl}).
  \end{lemm}

{\it Proof of the Lemma.} We denote by $ \Phi (h)$ the right hand side  and by
$ \Psi (h)$ the left hand side of  (\ref{series-form}).
Since  $F$ belongs to $S(H^2, Q)$, then we have (\ref{heat}), where
 the series is convergent and defines an element of  $S(H^2, Q)$.
Looking at the powers of  $h$, we are led to compare the coefficient of $h^k$
in $ \Phi (h)$
$$ a_k = \sum _{s+ \ell = k} \frac {1} {4^{\ell} \ell !} \Delta^{\ell} C_s^{weyl} (F , G)$$
with the coefficient of $h^k$ in $ \Psi (h)$
$$ b_k = \sum _{s+ \ell+ u = k}\frac {1} {4^{\ell + s } s!  \ell !}  C_u^{wick}
( \Delta ^{\ell } F , \Delta ^{s } G).$$
Let $\sigma ( \nabla _1, \nabla _2)$  be the differential operator on $H^2 \times H^2$ defined in (\ref{sigma}).
We denote by $R$ the restriction operator defined for all functions $F$ on $H^2 \times H^2$ by:
 $$ (RF) (X) = F(X , X).$$
We have
 $$ \Delta R = R \widetilde \Delta\qquad\qquad \widetilde \Delta = \Delta_1  +  \Delta_2 +
 2 \nabla _1 \cdot \nabla _2 \qquad\qquad \nabla _1 \cdot \nabla _2 = \sum _j \frac {\partial ^2} {\partial x_j \partial y_j}
 + \frac {\partial ^2} {\partial \xi_j \partial \eta_j}. $$
 We  then have to compare the following coefficients (operators acting on functions defined on  $H^2 \times H^2$)
 $$ A_k = \sum _{s+ \ell = k} \frac {1} { (2i)^s 4^{\ell}
 s! \ell !} (\widetilde \Delta )^{\ell}
 ( \sigma (\nabla _1 , \nabla _2))^s $$
$$ B_k = \sum _{s+ \ell + n = k}
\frac {1} { 2^n 4^{\ell + s}  s! \ell !}
\sum _{|\alpha | = n} \frac {1} { \alpha !}
(\partial _x - i \partial _{\xi} )^{\alpha}
(\partial _y + i \partial _{\eta} )^{\alpha} \Delta _1 ^{\ell}
\Delta _2 ^s. $$
Expanding $(\widetilde \Delta )^{\ell}$ using the binomial formula,
we are led to compare:
$$ A'_k = \sum _{\ell = 0}^k \sum _{u + s \leq \ell }
\frac {i ^{\ell - k}}  { 2 ^{k+u+s} (k-\ell )! u! s! (\ell - u -s)!}
(\nabla _1 \cdot \nabla _2) ^{\ell - u -s}
(\sigma (\nabla _1 , \nabla _2))^{k-\ell}
\Delta _1^u \Delta _2^s$$
$$ B'_k = \sum _{s+u + |\alpha| = k} \frac {1}
{2^{k+u+s} u! s! \alpha !} (\partial _x - i \partial _{\xi})^{\alpha}
(\partial _y + i \partial _{\eta})^{\alpha} \Delta _1^u \Delta _2^s. $$
Thus, we have to compare, for $t = u +s$ being integers smaller or equal than $k$:
$$ A_k (t) = \sum _{\ell = t}^k \frac {i ^{\ell - k}}
{ (k- \ell )! (\ell - t)!}
(\nabla _1 \cdot \nabla _2) ^{\ell - t}
(\sigma (\nabla _1 , \nabla _2))^{k-\ell}  $$
$$ B_k(t) =  \sum _{ |\alpha| = k-t }\frac {1} {\alpha !}
(\partial _x - i \partial _{\xi})^{\alpha}
(\partial _y + i \partial _{\eta})^{\alpha}.$$

The term $B_k$ is directly expanded. The term $A_k$ is less direct and contains the terms

$ \partial_x^a \partial_y^b \partial^{\alpha -a}_{\xi}\partial^{\alpha-b}_{\eta}$,
for  $|\alpha|= k-t$, with the  coefficient
$$
\sum_{c} \frac{i^{t+ |a+b+c-\alpha|+|c|}(-1)^{\alpha-b-c}}{(a+b+c-\alpha)!c!(\alpha-b-c)!(\alpha-a-c)!}
$$
the sum running over all  $c$ such that all the  multi-indices in the denominator
have positive components. A standard fact (concerning the hypergeometric distribution) then allows to conclude that $A_k(t)=B_k(t)$.

laurent.amour@univ-reims.fr\newline
LMR CNRS FRE 2011, Universit\'e de Reims Champagne-Ardenne,
 Moulin de la Housse, BP 1039,
 51687 REIMS Cedex 2, France.

lisette.jager@univ-reims.fr\newline
LMR CNRS FRE 2011, Universit\'e de Reims Champagne-Ardenne,
 Moulin de la Housse, BP 1039,
 51687 REIMS Cedex 2, France.

jean.nourrigat@univ-reims.fr\newline
LMR CNRS FRE 2011, Universit\'e de Reims Champagne-Ardenne,
 Moulin de la Housse, BP 1039,
 51687 REIMS Cedex 2, France.

\end{document}